\documentclass[12pt]{article}
\usepackage{amssymb,amsmath}
\usepackage[T2A]{fontenc} 
\usepackage[cp1251]{inputenc} 
\newtheorem{theorem}{Theorem}

\begin{document}

\begin{center}
{\bf LAX REPRESENTATION OF WZNW-LIKE SYSTEMS}
\end{center}
\begin{center}
A.V. Balandin  \footnote{corresponding author} \\
Department of Mathematics and Mechanics \\
N.I.Lobatchevsky Nizhny Novgorod State University\\
23 Gagarin ave., 603600 Nizhny Novgorod, Russia    \\
e-mail: balandin @ mm.unn.ru
\end{center}

\begin{center}
O.N. Pakhareva \\
Department of Mathematics and Mechanics\\
N.I.Lobatchevsky Nizhny Novgorod State University\\
23 Gagarin ave., 603600 Nizhny Novgorod, Russia \\
e-mail: pakhareva@rambler.ru
\end{center}

\vspace*{3mm}

The Lax representation and Backlund transformations for the systems
similar to WZNW (Wess-Zumino-Novicov-Witten) systems and non-abelian
affine Toda models are obtained in present paper. One of these
systems is a new integrable extension of well known sine-Gordon
equation.

\vspace*{3mm}

\section{Introduction}
Chiral-type systems (see e.g. \cite{Mesh})  are the systems of
partial differential equations of the form
\begin{equation} \label{S1} U^{\alpha}_{xy} +
G^{\alpha}_{\beta\gamma}U^{\beta}_xU^{\gamma}_y +Q^{\alpha} = 0.
\end{equation}
Here the Greek indices $\alpha ,\beta ,\gamma$ range from 1 to $n$
and the subscripts denote partial derivatives with respect to the
independent variables $x$ and $y$. The coefficients
$G^{\alpha}_{\beta\gamma},Q^{\alpha}$ are assumed to be smooth
functions of the variables  $U^1, U^2, ..., U^n$. The summation rule
over the repeated indices is also assumed.

If system (\ref{S1}) is the system of Euler-Lagrange equations
then the Lagrangian can be given by
\begin{equation}
\label{L1} L=g_{\alpha\beta}(U^{\delta})U^{\alpha}_x U^{\beta}_y +
a_{\alpha\beta}(U^{\delta})U^{\alpha}_x U^{\beta}_y+Q(U^{\delta}),
\end{equation}
where  $g_{\alpha\beta}$ is a non-degenerate symmetric matrix,
$a_{\alpha\beta}$ is a skew-symmetric matrix and $Q$ is a smooth
function of the variables $U^1, U^2, ..., U^n$. Notice that
Euler-Lagrange systems for Lagrangian (\ref{L1}) usually are
referred as general nonlinear sigma-models (see e.g. \cite{1a}
,\cite{Bilal}). These systems contain the special class of systems
referred as WZNW(Wess-Zumino-Novicov-Witten)(\cite{WZnw}
,\cite{wzNw} , \cite{wznW}, \cite{WZW_G}).

Recall that the WZNW systems can be defined  in the following way.
Let $H$ be a semi-simple Lie group, $(U^{\alpha})$  a local
coordinate system on $H$ and $\Phi^{\alpha}=T^{\alpha}_{\beta}
dU^{\beta}$  a basis of left-invariant forms. Then
\begin{equation}
\label{g2} G^{\alpha}_{\beta\gamma}=\tilde{T}^{\alpha}_{\delta}
T^{\delta}_{\beta , \gamma}
\end{equation}
are the coefficients of the first flat affine connection of the
group $H$. Here and further  $\tilde{T}^{\alpha}_{\beta}$ denotes
the inverse matrix for $T^{\alpha}_{\beta}$ and comma denotes the
partial derivative
$$T^{\delta}_{\beta ,\gamma}=\frac{\partial T^{\delta}_{\beta }}{\partial
U^{\gamma}}.$$ WZNW system associated with the semi-simple Lie group
$H$ is the system of the form (\ref{S1}) with $Q^{\alpha}=0$ and
coefficients $G^{\alpha}_{\beta\gamma}$ defined by (\ref{g2}).

To obtain the Lagrangian $L_1$ for WZNW system define a 2-form
$\sigma=a_{\alpha\beta}dU^{\alpha} \wedge dU^{\beta}$ which
satisfies the condition
$$d\sigma=\frac{2}{3}h^0_{\alpha\delta}C^{\delta}_{\beta\gamma}\Phi^{\alpha}
\wedge\Phi^{\gamma}\wedge\Phi^{\beta}.$$  Here
$C^{\alpha}_{\beta\gamma}$ are the structure constants of the group
$H$ determined by the basis $\Phi^{\alpha}$ and $h^0_{\alpha\beta}$
is the Killing form of the Lie algebra $\mathfrak{h}$  of the group
$H$ with respect to the dual basis. Such 2-form $\sigma$ exist, at
least locally, since 3-form $\Psi =\frac{2}{3} h^0_
{\sigma\delta}C_{\varphi\psi}^{\sigma}T^{\delta}_{\alpha}T^{\varphi}_{\beta}
T^{\psi}_{\gamma}dU^{\alpha}\wedge dU^{\gamma}\wedge dU^{\beta}$ is
closed due to the Jacobi identity. Then the Lagrangian $L_1$ is of
the form
\begin{equation}
\label{L3} L_1=h^0_{\gamma\delta}T_{\alpha}^{\gamma}
T_{\beta}^{\delta} U^{\alpha}_x U^{\beta}_y +
a_{\alpha\beta}U^{\alpha}_x U^{\beta}_y.
\end{equation}
It is well known that WZNW systems admit the Lax representation.

In Section 2  the Euler-Lagrange systems for the Lagrangian
$$L_2=h^0_{\gamma\delta}T^{\gamma}_{\alpha} T^{\delta}_{\beta }
U^{\gamma}_x U^{\delta}_y + k a_{\alpha\beta}U^{\alpha}_x
U^{\beta}_y$$ are considered. Here k is an arbitrary constant,
$h^0_{\alpha\beta}, T^{\alpha}_{\beta}, a_{\alpha\beta}$ are of the
same sense as above. We will call such systems the WZNW-like
systems. The Lax representation and Backlund transformation for
these WZNW-like systems are constructed in Section 2.

In Section 3 we give a slight addition to  Lesnov-Saveliev
construction \cite{LS} of Lax representation of nonlinear systems.
Lesnov-Saveliev method affords  to associate  some integrable system
with  symmetric space $G/H$. The systems obtained by this way are
usually referred as non-abelian affine Toda models
 ( \cite{Bilal} , \cite{NATM}).
 Our observation affords to obtain additional integrable
systems in case  $H$ is a direct product of simple Lie groups.
 Further we mention some new integrable systems associated with
 the symmetric spaces $SO(p+3)/(SO(p)\times SO(3)), SO(p+2)/(SO(p)\times SO(2)).$


\section{WZNW-like systems}

A construction of the Lax representation for some chiral field
systems
\begin{equation}
\label{m1}
U^{\alpha}_{xy} + G^{\alpha}_{\beta\gamma}U^{\beta}_xU^{\gamma}_y =
0
\end{equation}
is proposed here.

Let $\Theta^{a}$ be a basis of left-invariant forms of a Lie group
G. It is well known that the forms $\Theta ^{a}$ satisfy the Cartan
structure equations
\begin{equation}
\label{r2} d\Theta^{a}=C^{a}_{bc}\Theta^{c} \wedge \Theta^{b},
\end{equation}
where  $C^{a}_{bc} $ are the structure constants.

We say that system (\ref{m1}) admits a Lax representation with the
Lie  group G if the system
\begin{equation}
\label{r3} \Theta^{a}=A^{a} dx + B^{a} dy
\end{equation}
is completely integrable (in the sense of Frobenius) on solutions of
the system (\ref{m1}), i.e. if the substitution of expressions
(\ref{r3}) into (\ref{r2}) results  in the system equivalent to
(\ref{m1}).

We will consider the groups for which  Cartan's structure equations
can be written in the form
\begin{equation}
\label{m2} d\theta ^{\alpha}=D^{\alpha}_{\beta i}\varphi^i \wedge
\theta^{\beta},
\end{equation}
\begin{equation}
\label{m3} d\varphi^i=C^i_{jk}\varphi^k \wedge \varphi^j +
R^i_{\beta\gamma} \theta^{\gamma} \wedge \theta^{\beta},
\end{equation}
where  $D^{\alpha}_{\beta i},C^i_{jk},R^i_{\beta\gamma}=const$, the
Greece indices  $\alpha ,\beta ,\gamma$ range from 1 to $n$
(n=dim{G/H}), indices $i,j,k$ range from  1 to $r$.

It is well known \cite{Helg} that Eqs.(\ref{m2}) and (\ref{m3})
could also be considered as the structure equations of some locally
symmetric space $G/H$, where $C^{i}_{jk}$  are the structure
constants of the isotropy group $H$.

We look for the Lax representation of system (\ref{m1}) in the form
\begin{equation}
\label{m4} 
\theta ^{\alpha}=U_{\beta}^{\alpha}(\lambda U^{\beta}_x
dx+\frac{1}{\lambda} U^{\beta}_y dy),
\end{equation}
\begin{equation}
\label{m5} 
\varphi^i=(-\Gamma^i_{\alpha}+H^i_{\alpha})U^{\alpha}_{\beta}
U^{\beta}_xdx+(-\Gamma^i_{\alpha}-H^i_{\alpha})U^{\alpha}_{\beta}
U^{\beta}_ydy,
\end{equation}
where $\lambda$ is a parameter, $U^{\alpha}_{\beta},
\Gamma^i_{\alpha}, H^i_{\alpha}$ are the smooth functions of
variables $U^1, U^2, ... ,U^n$. Substitute  the expressions
(\ref{m4}),(\ref{m5}) into Eqs.(\ref{m2}). Taking into account
(\ref{m1}) and collecting similar terms,
we arrive at equations
\begin{equation}
\label{m6}U^{\alpha}_{\beta ,\gamma}-U^{\alpha}_{\delta}
G^{\delta}_{\beta\gamma}+ D^{\alpha}_{\mu
i}(\Gamma^i_{\nu}+H^i_{\nu})U^{\mu}_{\beta}U^{\nu}_{\gamma}=0,
\end{equation}
\begin{equation}
\label{m7} U^{\alpha}_{\gamma , \beta}-U^{\alpha}_{\delta}
G^{\delta}_{\beta\gamma}+ D^{\alpha}_{\mu
i}(\Gamma^i_{\nu}-H^i_{\nu})U^{\mu}_{\gamma}U^{\nu}_{\beta}=0.
\end{equation}
Subtracting  (\ref{m7}) from (\ref{m6}) and separating out the
symmetric and skew-symmetric parts of result, we obtain the
equalities
\begin{equation}
\label{m8} U^{\alpha}_{[\beta , \gamma]}+D^{\alpha}_{\mu
i}\Gamma^i_{\nu}U^{\mu}_{[\beta}U^{\nu}_{\gamma]}=0,
\end{equation}
\begin{equation}
\label{m9} D^{\alpha}_{\mu
i}H^i_{\nu}U^{\mu}_{(\beta}U^{\nu}_{\gamma)}=0. 
\end{equation}
Now from Eqs.(\ref{m6}),(\ref{m7}) in terms of (\ref{m8}),(\ref{m9})
it follows that
\begin{equation}
\label{m11} U^{\alpha}_{\delta}
G^{\delta}_{[\beta\gamma]}=D^{\alpha}_{\mu i}
H^i_{\nu}U^{\mu}_{[\beta}U^{\nu}_{\gamma]},
\end{equation}
\begin{equation}
\label{m12}
U^{\alpha}_{\delta}G^{\delta}_{(\beta\gamma)}=U^{\alpha}_{(\beta
,\gamma)} +D^{\alpha}_{\mu i} \Gamma
^i_{\nu}U^{\mu}_{(\beta}U^{\nu}_{\gamma)}.
\end{equation}
Analogously, substituting (\ref{m4}) and (\ref{m5}) into
Eqs.(\ref{m3}) and taking into account
(\ref{m1}),(\ref{m8}),(\ref{m9}),(\ref{m11}),(\ref{m12}), we get
\begin{equation}
\label{m13} \Gamma^i_{[\mu ; \nu ]}+H^i_{\sigma}D^{\sigma }_{[\mu
|j|}H^j_{\nu ]}-\Gamma^i_{\sigma} D^{\sigma}_{[\mu
|j|}\Gamma^j_{\nu ]}=C^i_{jk}\Gamma^k_{[\mu}\Gamma^j_{\nu
]}-C^i_{jk}H^k_{[\mu}H^j_{\nu ]}-R^i_{\mu \nu},
\end{equation}
\begin{equation}
\label{m14} H^i_{(\mu ; \nu )}=H^i_{\sigma}D^{\sigma}_{(\mu
|j|}\Gamma^j_{\nu )}+2C^i_{jk}H^k_{(\mu}\Gamma^j_{\nu )}.
\end{equation}
Here we assume $d\Gamma^i_{\nu}=\Gamma^i_{\nu ;
\sigma}U^{\sigma}_{\beta}dU^{\beta} ,$ $dH^i_{\nu}=H^i_{\nu ;
\sigma}U^{\sigma}_{\beta}dU^{\beta} $. So the following theorem is
proved.
\begin{theorem}
Let G be a Lie group with the structure equations
\eqref{m2},\eqref{m3} and $G^{\alpha}_{\beta\gamma},
U^{\alpha}_{\beta}, \Gamma^i_{\alpha}, H^i_{\alpha}$ are smooth
functions satisfying the conditions \eqref{m8}-\eqref{m14}. Then
equalities  \eqref{m4},\eqref{m5} define the Lax representation
 of the system \eqref{m1}.
\end{theorem}

\noindent {\bf Example 1.}

The proposed construction can be used to obtain well known  Lax
representation of chiral fields equations valued in symmetric space
$G/H$ (see e.g. \cite{Zah}).

Let $\theta^{\alpha}=U^{\alpha}_{\beta}dU^{\beta}$ are the basic
forms and $
\varphi^i=-\Gamma^i_{\alpha}U^{\alpha}_{\beta}dU^{\beta}$ are the
connection forms of the symmetric space $G/H$. Then from
(\ref{m2}),(\ref{m3}) it follows that
$$ U^{\alpha}_{[\mu ,
\nu]}=-D^{\alpha}_{\varphi
i}\Gamma^i_{\psi}U^{\varphi}_{[\mu}U^{\psi}_{\nu ]},$$
$$\Gamma^i_{[\mu , \nu ]}-\Gamma^i_{\sigma}D^{\sigma}_{[\mu |i|}
 \Gamma^i_{\nu ]}= C^i_{jk}\Gamma^k_{\mu}\Gamma^j_{\nu}-R^i_{\mu\nu}.$$
Assume  $H^i_{\alpha}=0$ in (\ref{m5}). Then the conditions
(\ref{m8}),(\ref{m9}),(\ref{m13}),(\ref{m14}) are fulfilled and
equations(\ref{m2})-(\ref{m5}) define the Lax representation of the
system (\ref{m1}), where
$$G^{\alpha}_{[\beta\gamma]}=0,$$
$$G^{\alpha}_{(\beta \gamma
)}=\tilde{U}^{\alpha}_{\sigma}[U^{\sigma}_{(\beta , \gamma
)}+D^{\alpha}_{\varphi
i}\Gamma^i_{\psi}U^{\varphi}_{(\beta}U^{\psi}_{\gamma )}],$$
i.e. $G^{\alpha}_{\beta\gamma}$ are the coefficients of the
canonical affine connection of the symmetric space $G/H$ with
respect to the holonomic basis.

\vspace*{3mm}
 \noindent {\bf Example 2.}

Let $H$ be a Lie group. Consider the symmetric space $(H \times
 H)/H$ with the canonical affine connection. Then the structure
 equations of the symmetric space $(H \times
 H)/H$ are of the form
 (\ref{m2}),(\ref{m3}), where $C^{i}_{jk}$ are the
 structure constants of the group $H$,
$D^{\alpha}_{\beta i}=2C^{\alpha}_{\beta i},\; R^i_{\alpha
\beta}=-C^i_{\alpha \beta}$ and all indices range from 1 to
$n=dimH$. Assuming $\Gamma^i_{\alpha}=0,
H^i_{\alpha}=\delta^i_{\alpha}$ in (\ref{m4}),(\ref{m5}), we obtain
the Lax representation of the system (\ref{m1}) with coefficients
$G^{\alpha}_{\beta\gamma}$ of the form
$$G^{\alpha}_{[\beta\gamma]}=2\tilde{U}^{\alpha}_{\delta}
C^{\delta}_{\varphi\psi}U^{\varphi}_{\beta}U^{\psi}_{\gamma},$$
$$G^{\alpha}_{(\beta\gamma)}=2\tilde{U}^{\alpha}_{\delta}U^{\delta}_{\beta ,\gamma}.$$
Here $U^{\alpha}_{\beta}$ are the smooth functions with the
following properties $$U^{\alpha}_{[\beta , \gamma]}=0,$$
$$det||U^{\alpha}_{\beta}||\neq 0.$$

\vspace*{5mm}
\begin{theorem}
Let $H$ be a semi-simple Lie group and

\noindent 1) $(U^{\alpha})$ is a local coordinate system on the
group  $H$;

\noindent 2) $C^{\alpha}_{\beta\gamma}$ are the structure constants
of the Lie group $H$ with respect to the basis of left-invariant
forms $ \Phi^{\alpha}=T^{\alpha}_{\beta}dU^{\beta};$

\noindent 3) $h^0_{\alpha\beta}$ is the Killing form of the Lie
algebra $\mathfrak{h}$ of the group $H$ given in the dual basis;

\noindent 4) $h_{\alpha\beta}$ is the Killing metric of the group
$H,$ i.e.
\begin{equation}
\label{mm1} h_{\alpha\beta}=h^0_{\varphi\psi }T^{\varphi
}_{\alpha}T^{ \psi}_{\beta};
\end{equation}

\noindent 5) $\sigma=a_{\alpha\beta}dU^{\alpha}\wedge dU^{\beta}$ is
the 2-form which satisfies the condition
$$d\sigma=\frac{2}{3}h^0_{\delta\sigma}C^{\sigma}_{\beta\gamma}\Phi^{\delta}
\wedge\Phi^{\gamma}\wedge\Phi^{\beta}.$$

Then the system of  Euler-Lagrange equations for  Lagrangian
\begin{equation}
\label{m16} 
L=h_{\alpha\beta}U^{\alpha}_x U^{\beta}_y +
ka_{\alpha\beta}U^{\alpha}_x U^{\beta}_y
\end{equation}
admits the Lax representation. Here $k\neq \pm 1$ is an arbitrary
constant.
\end{theorem}

{\bf Proof.} Consider the symmetric space $(H\times H)/H$ with the
structure equations (\ref{m2}),(\ref{m3}), where $D^{\alpha}_{\beta
i}=2C^{\alpha}_{\beta i},\; R^i_{\alpha \beta}=-C^i_{\alpha \beta}$
and all indices range from 1 to $n$. Assume
$$U^{\alpha}_{\beta}=\frac{1}{2}\sqrt{k^2-1}T^{\alpha}_{\beta},$$
$$\Gamma^i_{\alpha}=
-\frac{1}{\sqrt{k^2-1}}\delta^i_{\alpha},
H^i_{\alpha}=\frac{k}{\sqrt{k^2-1}}\delta^i_{\alpha},$$ where $k\neq
\pm 1$ is an arbitrary constant. Then the expressions
(\ref{m4}),(\ref{m5}) define the Lax representation for the system
\begin{equation}
\label{m15} 
U^{\alpha}_{xy}+\tilde{T}^{\alpha}_{\sigma}[T^{\sigma}_{(\beta
,\gamma)}+kC^{\sigma} _{\varphi\psi}
T^{\varphi}_{\beta}T^{\psi}_{\gamma}]U^{\beta}_x U^{\gamma}_y=0.
\end{equation}

Now prove that the system (\ref{m15}) is the system of Euler's
equations for Lagrangian (\ref{m16}). Indeed, the Euler equations
for the Lagrangian (\ref{m16}) results in the system  (\ref{m1}),
where
\begin{equation}
\label{m17} 
G^{\alpha}_{\beta\gamma} = \frac{1}{2}\tilde{h}^{\alpha\delta}
(h_{\delta\gamma,\beta} + h_{\delta\beta,\gamma} -
h_{\beta\gamma,\delta}) +
\frac{3}{2}\tilde{h}^{\alpha\delta}a_{[\delta\gamma,\beta]}.
\end{equation}
One can easily verify that in our case we have the equalities
\begin{equation}
\label{m17a} 
G^{\alpha}_{(\beta\gamma)} = \frac{1}{2}\tilde{h}^{\alpha\delta}
(h_{\delta\gamma,\beta} + h_{\delta\beta,\gamma} -
h_{\beta\gamma,\delta})
=\tilde{T}^{\alpha}_{\delta}T^{\delta}_{(\beta,\gamma)}.
\end{equation}
So the coefficients  $G^{\alpha}_{\beta\gamma}$ of the Euler
equations are of the form (\ref{m15}) if the torsion
\begin{equation}
\label{m17b} 
A^{\alpha}_{\beta\gamma} =G^{\alpha}_{[\beta\gamma ]}=
k\tilde{T}^{\alpha}_{\sigma}C^{\sigma} _{\varphi\psi}
T^{\varphi}_{\beta}T^{\psi}_{\gamma}
\end{equation}
of the connection satisfies the following condition $
A_{\alpha\beta\gamma} = h_{\delta\alpha}A_{\beta\gamma}^{\delta} =
\frac{3}{2}a_{[\alpha\gamma ,\beta]}.$ Taking into account
(\ref{mm1}), we obtain  $A_{\alpha\beta\gamma} =
kh^0_{\sigma\delta}C_{\varphi\psi}^{\sigma}T^{\delta}_{\alpha}T^{\varphi}
_{\beta}T^{\psi}_{\gamma}.$

Notice that 3-form $\Psi =\frac{2}{3} h^0_
{\sigma\delta}C_{\varphi\psi}^{\sigma}T^{\delta}_{\alpha}T^{\varphi}_{\beta}
T^{\psi}_{\gamma}dU^{\alpha}\wedge dU^{\gamma}\wedge dU^{\beta}$ is
closed due to the Jacobi identity. Therefore there exist, at least
locally, a   2-form $\sigma$ such that the equality $d\sigma = \Psi$
is fulfilled. So assuming $ a_{\alpha\beta}dU^{\alpha}\wedge
dU^{\beta} = \sigma$, we obtain that the Euler equations for
Lagrangian (\ref{m16}) coincide with the system (\ref{m15}).
$\square$

\vspace*{3mm} \noindent {\bf Remark.}  One can easily verify that
the coefficients $G^{\alpha}_{\beta\gamma }$ defined by equalities
(\ref{m17a}),(\ref{m17b}) are the coefficients of the affine
connection of homogeneous reductive space which can be characterized
by the following conditions:
$$\nabla R^{\alpha}_{\beta\gamma\delta} = 0,\; \nabla A^{\alpha}_{\beta\gamma} = 0.$$
Moreover, one can see that  the Ricci identities
$$R^{\alpha}_{[\beta\gamma\delta]}=0$$
are true for the connection with coefficients defined by
(\ref{m17a}),(\ref{m17b}).

Notice that the Lax representation without parameter for the systems
with such connection was mentioned in \cite{Balandin}.

\vspace*{3mm} The Backlund transformation for the systems
(\ref{m15}) can be obtained from the following theorem

\begin{theorem}
Let
$\overline{C}^{\alpha}_{\beta\gamma},\widehat{C}^{\alpha}_{\beta\gamma}$
be the structure constants of n-dimensional Lie algebras and
functions  $U^{\alpha}_{\beta}=U^{\alpha}_{\beta}(U^{\gamma}),
V^{\alpha}_{\beta}=V^{\alpha}_{\beta}(V^{\gamma})$ satisfy the
equations:
\begin{equation}
\label{b1} U^{\alpha}_{[\beta ,
\gamma]}=\overline{C}^{\alpha}_{\varphi\psi}U^{\varphi}_{\beta}U^{\psi}_{\gamma},\;\;
det||U^{\alpha}_{\beta}|| \neq 0,
\end{equation}
\begin{equation}
\label{b2} V^{\alpha}_{[\beta ,
\gamma]}=\widehat{C}^{\alpha}_{\varphi\psi}V^{\varphi}_{\beta}V^{\psi}_{\gamma},\;\;
det||V^{\alpha}_{\beta}|| \neq 0,
\end{equation}
Then the equalities
\begin{equation}
\label{b3}
U^{\alpha}_{\beta}U^{\beta}_x=V^{\alpha}_{\beta}V^{\beta}_x,
\end{equation}
\begin{equation}
\label{b4}
U^{\alpha}_{\beta}U^{\beta}_y=-V^{\alpha}_{\beta}V^{\beta}_y
\end{equation}
define the Backlund transformations between the  systems
\begin{equation} \label{s1}
U^{\alpha}_{xy}+\tilde{U}^{\alpha}_{\delta}[U^{\delta}_{(\beta
,\gamma)}+\widehat{C}^{\delta} _{\varphi\psi}
U^{\varphi}_{\beta}U^{\psi}_{\gamma}]U^{\beta}_x U^{\gamma}_y=0,
\end{equation}
\begin{equation}
\label{s2}
V^{\alpha}_{xy}+\tilde{V}^{\alpha}_{\delta}[V^{\delta}_{(\beta
,\gamma)}+\overline{C}^{\delta} _{\varphi\psi}
V^{\varphi}_{\beta}V^{\psi}_{\gamma}]V^{\beta}_x V^{\gamma}_y=0
\end{equation}
\end{theorem}

{\bf Proof.} Indeed, take the derivatives of equalities
(\ref{b3}),(\ref{b4}) with respect to variables  $x$ and $y$. Then
adding and subtracting  the obtained equalities and taking into
account (\ref{b1}),(\ref{b2}),(\ref{b3}),(\ref{b4}), we arrive at
the systems (\ref{s1}),(\ref{s2}). $\square$ \vspace*{3mm}

{\bf Corollary 1.} {\it 
The equalities \eqref{b3},\eqref{b4} define the Backlund
transformations between the  systems
\begin{equation}
\label{s1a}
U^{\alpha}_{xy}+\tilde{U}^{\alpha}_{\delta}[U^{\delta}_{(\beta
,\gamma)}+kC^{\delta} _{\varphi\psi}
U^{\varphi}_{\beta}U^{\psi}_{\gamma}]U^{\beta}_x U^{\gamma}_y=0,
\end{equation}
\begin{equation}
\label{s2a}
V^{\alpha}_{xy}+\tilde{W}^{\alpha}_{\delta}[W^{\delta}_{(\beta
,\gamma)}+\frac{1}{k}C^{\delta} _{\varphi\psi}
W^{\varphi}_{\beta}W^{\psi}_{\gamma}]V^{\beta}_x V^{\gamma}_y=0,
\end{equation}
were $k=const$ and functions
$U^{\alpha}_{\beta}=U^{\alpha}_{\beta}(U^{\gamma}),
W^{\alpha}_{\beta}=W^{\alpha}_{\beta}(V^{\gamma})$ satisfy the
equations
\begin{equation} \label{b2''} U^{\alpha}_{[\beta ,
\gamma]}=C^{\alpha}_{\varphi\psi}U^{\varphi}_{\beta}U^{\psi}_{\gamma},\;\;
det||U^{\alpha}_{\beta}|| \neq 0,
\end{equation}
\begin{equation} \label{b2'} W^{\alpha}_{[\beta ,
\gamma]}=C^{\alpha}_{\varphi\psi}W^{\varphi}_{\beta}W^{\psi}_{\gamma},\;\;
det||W^{\alpha}_{\beta}|| \neq 0.
\end{equation}}

{\bf Proof.} Assume
$\overline{C}^{\alpha}_{\beta\gamma}=C^{\alpha}_{\beta\gamma},
\widehat{C}^{\alpha}_{\beta\gamma}=kC^{\alpha}_{\beta\gamma},
W^{\alpha}_{\beta}=kV^{\alpha}_{\beta}.$ Then the functions
$U^{\alpha}_{\beta}$,  $W^{\alpha}_{\beta}$ satisfy the
Eqs.(\ref{b2'}),(\ref{b2''}) and the systems (\ref{s1}),(\ref{s2})
coincide with the systems (\ref{s1a}),(\ref{s2a}).$\square$



\section{Systems associated with the symmetric space $G/(H_1 \times ... \times H_p)$ }

We use the local coordinate approach to Lesnov-Saveliev construction
in order to find the Lax representation for some systems of the form
(\ref{S1}).

Let $G/H$ be a symmetric space with the structure equations of the
form
\begin{equation}
\label{m22} d\omega^{\alpha '}= D^{\alpha '}_{\beta ' \gamma}
\theta^{\gamma} \wedge\omega^{\gamma '},
\end{equation}
\begin{equation}
\label{m23} d\theta^{\alpha} =
C^{\alpha}_{\beta\gamma}\theta^{\gamma} \wedge \theta^{\beta} +
R^{\alpha}_{\beta ' \gamma '} \omega^{\gamma '} \wedge
\omega^{\beta '}.
\end{equation}
Here the Greek  indices $\alpha ,\beta ,\gamma$ range from 1 to
$n$ and indices $\alpha ',\beta ',\gamma '$ range from $n+1$ to
$r$. We assume
\begin{equation}
\label{m24} \omega^{\alpha '}  = \lambda M^{\alpha '}dx
+\frac{1}{\lambda}N^{\alpha '}dy,
\end{equation}
\begin{equation}
\label{m25} \theta^{\alpha} = T^{\;\alpha}_{1\beta}U^{\beta}_x dx
+T^{\;\alpha}_{2\beta}U^{\beta}_y dy,
\end{equation}
where  $M^{\alpha '}, N^{\alpha '},T^{\;\alpha }_{1 \beta},
T^{\;\alpha}_{2\beta}$ are the smooth functions of variables
$U^1,...,U^n$.One can easily verify that  the following theorem is
true  \cite{phl01}.

\begin{theorem}
Let  $C^{\alpha}_{\beta\gamma}$ be the structure constants of the
isotropy group $H$ of some locally symmetric space  $G/H$  with the
structure equations \eqref{m22},\eqref{m23}. Assume that there exist
matrices $\|T^{\;\alpha}_{1\beta}\|,\|T^{\;\alpha}_{2\beta}\|$ and
functions $M^{ \alpha '},N^{\alpha '}$ with the following
properties:
\begin{equation}
\label{m26} T^{\;\alpha}_{i[\beta ,\gamma]} =
C^{\alpha}_{\mu\nu}T^{\;\mu}_{i\beta}T^{\;\nu}_{i\gamma}\;
(i=1,2),\;\;
\end{equation}
\begin{equation}
\label{m27} det
\|T^{\;\alpha}_{1\beta}-T^{\;\alpha}_{2\beta}\|\neq0,
\end{equation}
\begin{equation}
\label{m28} M^{\alpha '}_{,\delta} = D^{\alpha '}_{\beta
'\gamma}T^{\;\gamma}_{2\delta}M^{\beta '},
\end{equation}
\begin{equation}
\label{m29} N^{\alpha '}_{,\delta}= D^{\alpha '}_{\beta '
\gamma}T^{\;\gamma}_{1\delta}N^{\beta '}.
\end{equation}
Then equalities \eqref{m24},\eqref{m25} define the Lax
representation of the system \eqref{S1} if the functions $G
^{\alpha}_{\beta\gamma},Q^{\alpha}$ are of the form
$$
G^{\alpha}_{\beta\gamma} =
\tilde{P}^{\alpha}_{\delta}[P^{\delta}_{(\beta ,\gamma)} +
2C^{\delta}_{\varphi\psi}S^{\varphi}_{\beta}S^{\psi}_{\gamma} -
2C^{\delta}_{\varphi\psi}P^{\varphi}_{(\gamma}S^{\psi}_{\beta)}],$$
$$Q^{\alpha} = - \tilde{P}^{\alpha}_{\delta}R^{\delta}_{\beta ' \gamma'}N^{\gamma '}M^{\beta '},$$
where  $S^{\alpha}_{\beta}=\frac{1}{2}(T^{\; \alpha}_{1\beta}+T^{\;
\alpha}_{2\beta}),\; P^{\alpha}_{\beta}= \frac{1}{2}(T^{\;
\alpha}_{1\beta}-T^{\; \alpha}_{2\beta})$ and $ \tilde{P}$ denotes
the inverse matrix for $P$.
\end{theorem}

\vspace*{3mm} The question under consideration is how to find the
functions $T^{\;\alpha}_{i \beta},M^{\alpha '},N^{\alpha '}$
satisfying the condition (\ref{m26})-(\ref{m29}). One solution can
be obtained by the following way.
Let  $C^{\alpha}_{\beta\gamma}$ be the structure constants of the
group  $H$ with respect to a basis of left-invariant forms $\Phi
^{\alpha}=T^{\alpha}_{\beta}dU^{\beta}$.
%
Assume  $ T^{\;\alpha}_{1\beta}=T^{\alpha}_{\beta},\;T^{\;
\alpha}_{2\beta}=0$ (or
$T^{\;\alpha}_{1\beta}=0,\;T^{\;\alpha}_{2\beta}=T^{\alpha}_{\beta}$).
Then one can see that (\ref{m26}),(\ref{m27})  and the compatibility
conditions for the systems (\ref{m28}),(\ref{m29}) are fulfilled. So
the equalities (\ref{m22})-(\ref{m25}) define the Lax representation
of the system (\ref{S1}) with the coefficients $Q^{\alpha}$ and
$G^{\alpha}_{\beta\gamma}$ mentioned in theorem 4. Notice that this
Lax representation coincide with one obtained by Lesnov-Saveliev
\cite{LS}. Further we give another solution of the equations
(\ref{m26})-(\ref{m29}) in case $H$ is a direct product  of simple
groups $H_1,...,H_p$.
\vspace*{3mm}

Let $G/H$ be a locally symmetric space and $H=H_1 \times H_2$, where
$H_1,H_2$ are the simple Lie groups. Split the local coordinates
$(U^{\alpha})$ into to groups $(U^{\alpha_1},U^{\alpha_2})$
according to the decomposition of the group $H$ and rewrite the
structure equations of the symmetric space $G/H$ in the form
\begin{equation}
\label{m42} d\omega^{\beta '}= D^{\beta '}_{\gamma '\alpha_1}
\theta^{\alpha_1} \wedge \omega^{ \gamma '}+D^{\beta '}_{\gamma '
\alpha_2} \theta^{\alpha_2} \wedge \omega^{\gamma '},
\end{equation}
\begin{equation}
\label{m43} d\theta^{\alpha_1} = C^{\alpha_1}_{\beta_1
\gamma_1}\theta^{\gamma_1} \wedge \theta^{\beta_1} +
R^{\alpha_1}_{\beta '\gamma '} \omega^{\gamma '} \wedge
\omega^{\beta '},
\end{equation}
\begin{equation}
\label{m44} d\theta^{\alpha_2} = C^{\alpha_2}_{\beta_2 \gamma_2
}\theta^{\gamma_2} \wedge \theta^{\beta_2} + R^{\alpha_2}_{\beta
'\gamma '} \omega^{\gamma '} \wedge \omega^{\beta '}.
\end{equation}
We will use the following notations.

\noindent 1)
$\mathfrak{g},\mathfrak{h},\mathfrak{h_1},\mathfrak{h_2}$ are the
Lie algebras of the groups Lie $G,H,H_1,H_2$ respectively.

\noindent 2) $C^{\alpha_1}_{\beta_1
\gamma_1},C^{\alpha_2}_{\beta_2 \gamma_2}$ are the structure
constants of the Lie groups $H_1,H_2$ with respect to the bases of
left-invariant forms
$\Phi^{\alpha_1}=T^{\alpha_1}_{\beta_1}dU^{\beta_1},\Phi^{\alpha_2}
=T^{\alpha_2}_{\beta_2 }dU^{\beta_2}.$

\noindent 3) $h^0_{\alpha_1 \beta_1},h^0_{\alpha_2 \beta_2}$ are
the matrices of the Killing forms $h^0_1(\cdot ,
\cdot),h^0_2(\cdot , \cdot)$ of the Lie algebras
$\mathfrak{h_1},\mathfrak{h_2}$ given in the bases dual to the
bases $\Phi^{\alpha_1},\Phi^{\alpha_2}$ and $g^0(\cdot , \cdot)$
is the Killing form of the algebra $\mathfrak{g}$.

\noindent 4) $S_i$ are the constants satisfying the equalities
\begin{equation}
\label{m34} h^0_i ( \cdot , \cdot)=S_ig^0( \cdot ,
\cdot)_{|_{\mathfrak{h_i}}}\;\;(i=1,2).
\end{equation}
Such constants exist due to simplicity of the subalgebras
$\mathfrak{h_1},\mathfrak{h_2}$.

\noindent 5) $h_{\alpha_1 \beta_1}, h_{\alpha_2 \beta_2}$ are the
Killing metrics of the groups  $H_1$ and $H_2$  respectively, i.e.
\begin{equation}
\label{m35} h_{\alpha_i \beta_i}=h^0_{\gamma_i
\delta_i}T^{\gamma_i}_{\alpha_i}T^{\delta_i}_{\beta_i},\;\;
(i=1,2).
\end{equation}

\noindent 6) $\sigma _i =a_{\alpha_i \beta_i}dU^{\alpha_i} \wedge
dU^{\beta_i}\;(i=1,2)$ are the 2-forms which satisfy the conditions
\begin{equation}
\label{m36} d \sigma_i=\frac{2}{3}h^0_{\alpha_i
\delta_i}C^{\delta_i}_{\beta_i \gamma_i}\Phi ^{\alpha_i} \wedge \Phi
^{\gamma_i} \wedge\Phi ^{\beta_i}.
\end{equation}

\begin{theorem}
Let $G/H$ be a locally symmetric space, $G$ the semi-simple Lie
group and $H=H_1\times H_2$, where $H_1,H_2$ are the simple Lie
groups.

Then the Euler's system for Lagrangian
$$L=S_2[h_{\alpha _1 \beta_1}(U^{\gamma _1})+\varepsilon _1 a_{\alpha_1
\beta_1}(U^{\gamma_1})]U^{\alpha_1}_x U^{\beta_1}_y+$$
\begin{equation} \label{mml1}
+S_1[h_{\alpha _2 \beta_2}(U^{\gamma _2})+\varepsilon _2
a_{\alpha_2 \beta_2}(U^{\gamma_2})]U^{\alpha_2}_x U^{\beta_2}_y +Q
\end{equation}
 admits
the Lax representation if
\begin{equation}
\label{mml1q} Q=4S_1 S_2 g^0_{\gamma '\delta '}M^{\gamma '}
N^{\delta '},
\end{equation}
$$\varepsilon _i =\pm 1 \;(i=1,2),$$
and functions $M^{\gamma '}, N^{\delta '}$ satisfy the following
conditions:

\begin{equation}
\label{m48} M^{\gamma
'}_{,\beta_1}=\frac{1-\varepsilon_1}{2}D^{\gamma '}_{\delta '
\alpha_1}T^{\alpha_1 }_{\beta_1}M^{\delta '},\;\; M^{\gamma
'}_{,\beta_2}=\frac{1-\varepsilon_2}{2}D^{\gamma '}_{\delta '
\alpha_2}T^{\alpha_2}_{\beta_2}M^{\delta '},
\end{equation}
\begin{equation}
\label{m49} N^{\gamma
'}_{,\beta_1}=\frac{1+\varepsilon_1}{2}D^{\gamma '}_{\delta '
\alpha_1}T^{\alpha_1 }_{\beta_1}N^{\delta '},\;\; N^{\gamma
'}_{,\beta_2}=\frac{1+\varepsilon_2}{2}D^{\gamma '}_{\delta
'\alpha_2}T^{\alpha_2}_{\beta_2}N^{\delta '}.
\end{equation}
\end{theorem}

 {\bf Proof.} Assume
\begin{equation} \label{m45}
\theta^{\alpha_1}=\frac{1+\varepsilon_1}{2}T^{\alpha_1
}_{\beta_1}U^{\beta_1}_x dx+\frac{1-\varepsilon_1}{2}T^{\alpha_1
}_{\beta_1}U^{\beta_1}_y dy,
\end{equation}
\begin{equation}
\label{m46} \theta^{\alpha_2}=\frac{1+\varepsilon_2}{2}T^{\alpha_2
}_{\beta_2}U^{\beta_2}_x dx+\frac{1-\varepsilon_2}{2}T^{\alpha_2
}_{\beta_2}U^{\beta_2}_y dy,
\end{equation}
\begin{equation}
\label{m47} \omega^{\gamma '}  = \lambda M^{\gamma '}dx
+\frac{1}{\lambda}N^{\gamma '}dy.
\end{equation}
Substituting the expressions (\ref{m45})-(\ref{m47}) into the
structure equations (\ref{m42})-(\ref{m44}), we arrive at the system
of the form (\ref{S1}) with coefficients
\begin{equation}
\label{mc1} G^{\alpha_1}_{\beta_1
\gamma_1}=\tilde{T}^{\alpha_1}_{\delta_1} (T^{\delta_1}_{(\beta_1
, \gamma_1)}+ \varepsilon_1 T^{\delta_1}_{[\beta_1 , \gamma_1]}),
\end{equation}
\begin{equation}
\label{mc2} G^{\alpha_2}_{\beta_2
\gamma_2}=\tilde{T}^{\alpha_2}_{\delta_2} (T^{\delta_2}_{(\beta_2 ,
\gamma_2)}+ \varepsilon_2 T^{\delta_2}_{[\beta_2 , \gamma_2]}),
\end{equation}
\begin{equation}
\label{mc3} Q^{\alpha_1}=2\varepsilon_1
\tilde{T}^{\alpha_1}_{\delta_1}R^{\delta_1}_{\beta ' \gamma
'}M^{\gamma '}N^{\beta '},
\end{equation}
\begin{equation}
\label{mc4} Q^{\alpha_2}=2\varepsilon_2
\tilde{T}^{\alpha_2}_{\delta_2}R^{\delta_2}_{\beta ' \gamma
'}M^{\gamma '}N^{\beta '}.
\end{equation}
Prove that this system is the Euler-Lagrange system for the
Lagrangian (\ref{mml1}). Indeed, the Euler equations for the
Lagrangian (\ref{mml1}) results in the system
$$U^{\alpha_1}_{xy}+ G^{\alpha_1}_{\beta_1 \gamma_1}U^{\beta_1}_xU^{\gamma_1}_y-
\frac{1}{2S_2}\tilde{h}^{\alpha_1 \beta_1}Q_{, \beta_1}=0,$$
$$U^{\alpha_2}_{xy}+ G^{\alpha_2}_{\beta_2 \gamma_2}U^{\beta_2}_xU^{\gamma_2}_y-
\frac{1}{2S_1}\tilde{h}^{\alpha_2 \beta_2}Q_{, \beta_2}=0,$$ where
the coefficients $G^{\alpha_1}_{\beta_1
\gamma_1},G^{\alpha_2}_{\beta_2 \gamma_2}$ are of the form
(\ref{mc1}),(\ref{mc2}). The proof of this fact is analogous to the
proof of the theorem 2.

Now prove  that the functions $ -\frac{1}{2S_2}\tilde{h}^{\alpha_1
\beta_1}Q_{, \beta_1}$ coincide with the functions (\ref{mc3}).
Let $\mathfrak{g}=\mathfrak{h}\oplus \mathfrak{m}$ be a canonical
decomposition, i.e.
$$[\mathfrak{h},\mathfrak{h}]\subset \mathfrak{h},
[\mathfrak{h},\mathfrak{m}] \subset \mathfrak{m},
[\mathfrak{m},\mathfrak{m}] \subset \mathfrak{h}.$$ Then due to
the properties of the Killing form and (\ref{m34})  it is true
that
\begin{equation}
\label{m40} h^0_1(h,[m_1,m_2])=S_1 g^0(m_1,[m_2,h])
\end{equation}
for any arbitrary elements $ h \in \mathfrak{h_1}, \;\;m_1,m_2 \in
\mathfrak{m}$.

The equalities (\ref{m40}) are equivalent to the following
conditions
\begin{equation}
\label{m41}
 h^0_{\alpha_1 \varphi_1}R^{\varphi_1}_{\beta '\gamma '}=S g^0_{\beta '\delta '}
 D^{\delta '}_{\gamma ' \alpha_1}.
\end{equation}
Now differentiating expression (\ref{mml1q}) and taking into account
(\ref{m48}),(\ref{m49}) and (\ref{m41}), we obtain
$$Q_{,\beta_1}=4S_1 S_2g^0_{\gamma ' \delta '}(M^{\gamma '}_{, \beta_1}N^{\delta '}+
M^{\gamma '}N^{\delta '}_{, \beta_1})=$$
$$=4S_1 S_2 g^0_{\gamma ' \delta '}(\frac{1-\varepsilon_1}{2}D^{\gamma '}_{\psi ' \alpha_1}
T^{\alpha_1}_{\beta_1}M^{\psi '}N^{\delta
'}+\frac{1+\varepsilon_1}{2}D^{\delta '}_{\psi '
\alpha_1}T^{\alpha_1}_{\beta_1}M^{\gamma '}N^{\psi '})=$$
$$=-4\varepsilon_1 S_2h^0_{\alpha_1 \varphi_1}R^{\varphi_1}
_{\delta ' \gamma '}T^{\alpha_1}_{\beta_1}M^{\gamma '}N^{
 \delta '}$$ and the functions $ -\frac{1}{2S_2}\tilde{h}^{\alpha_1
\beta_1}Q_{, \beta_1}$ coincide with the functions (\ref{mc3}).
Analogously one can verify that the  functions $
-\frac{1}{2S_1}\tilde{h}^{\alpha_2 \beta_2}Q_{, \beta_2}$ coincide
with the functions (\ref{mc4}).$\square$

\vspace*{3mm} \noindent {\bf Remark.} In case
$\varepsilon_1=\varepsilon_2$ we come to the system which can be
obtained by means of Lesnov-Saveliev construction.

\vspace*{3mm} Analogously can be proved the following theorem.

\begin{theorem}
Let $G/H$ be a locally symmetric space, $G$  the semi-simple Lie
group and $H=H_1\times ... \times H_p$, where $H_1,...,H_p$ are the
simple Lie groups.

Then the Euler's system for Lagrangian
$$L=\sum_{i=1}^{p} \frac{S}{S_i}[h_{\alpha _i \beta_i}(U^{\gamma _i})+\varepsilon _i
a_{\alpha_i \beta_i}(U^{\gamma_i})]U^{\alpha_i}_x U^{\beta_i}_y
+4Sg^0_{\beta '\gamma '}M^{\beta '} N^{\gamma '}$$ admits the Lax
representation. Here $\varepsilon _i =\pm 1$;  $h_{\alpha_i \beta_i}
$ are the Killing metrics of the groups $H_i$ $(i=\overline{1,p})$;
$S=S_1...S_p$; constants $S_i$ are defined by equalities
$$h^0_i( \cdot , \cdot)=S_ig^0( \cdot ,
\cdot)_{|_{\mathfrak{h}_i}};$$  $a_{\alpha_i \beta_i} $ are
defined as in previous theorems by 2-forms $\sigma_i;$ and
$M^{\beta '}, N^{\gamma '}$ are defined analogously to previous
theorem.
\end{theorem}

\vspace*{3mm}

\noindent {\bf Example 3.}(The system associated with
$SO(6)/(SO(3)\times SO(3)).$)

Choose the coordinates $U^1,U^2,U^3$ of the group $SO(3)$ to be
Euler's angles. Then the left-invariant forms can be written in the
following form
$$\Phi^1 = -\cos U^2 dU^3 - \sin
U^2 \sin U^3dU^1,$$
$$\Phi^2 = \sin U^3 \cos U^2 dU^1 - \sin U^2dU^3,\;$$
$$\Phi^3 = -\cos U^3 dU^1 - dU^2$$
and
$$d\Phi^1 = \Phi^3 \wedge \Phi^2,\;
\Phi^2 = \Phi^1 \wedge \Phi^3,\;\Phi^3 = \Phi^2 \wedge \Phi^1$$ are
the structure equations of the group  $SO(3).$ On the second group
$H_2=SO(3)$  we choose the local coordinates  $U^4,U^5,U^6$ and
left-invariant forms $\Phi^4,\Phi^5,\Phi^6$ analogously.

Rewrite the structure equations of the symmetric space
$SO(6)/(SO(3)\times SO(3))$ in the form
 \begin{equation}
\label{m51} d\Omega = \Omega \wedge  \Omega,\; \Omega = \left
\Vert
\begin{array}{cc}
\theta^{a}_{b}&\omega^{a}_{b'}\\
\omega^{a'}_{b}&\theta^{a'}_{b'}
\end{array}
\right \Vert ,
\end{equation}
where the forms $\theta^{a}_{b}=-\theta^{b}_{a},
\theta^{a'}_{b'}=-\theta^{b'}_{a'}, \omega^{a'}_{ b} = -
\omega_{a'}^{b}$, the  indices  $a,b,c...$ range from  1 to 3 and $a
', b',...$ range from 4 to 6.

Assume  $\varepsilon_1 =-1, \varepsilon_2 =1,$ i.e.
$$\theta^1_2=(-\cos U^3U^1_y-U^2_y)dy,\;
\theta^1_3=(-\sin U^3\cos U^2U^1_y+\sin U^2U^3_y)dy,$$
\begin{equation}
\label{m52} \theta^2_3=(-\cos U^2U^3_y-\sin U^2\sin U^3U^1_y)dy,\;
\theta^4_5=(-\cos U^6U^4_x-U^5_x)dx,
\end{equation}
$$
\theta^4_6=(-\sin U^6\cos U^5U^4_x+\sin U^5U^6_x)dx,\;
\theta^5_6=(-\cos U^5U^6_x-\sin U^5\sin U^6U^4_x)dx,
$$
\begin{equation}
\label{m53} \omega^{a}_{a'}=\lambda M^{a}_{a'}dx+
\frac{1}{\lambda}N^{a}_{a'}dy.
\end{equation}
Substituting the expressions (\ref{m52}),(\ref{m53}) into
Eqs.(\ref{m51}) and equating the coefficients with $\lambda$ and
$\frac{1}{\lambda}$ , we arrive at the system of partial
differential equations for the functions $M^{a}_{a '}, N^{a}_{a '}.$
Integrating this system, we find the simplest solution: $
M^1_{6}=\sin U^2\sin U^3,\;M^2_{6}=-\cos U^2\sin U^3,\; M^3_{6}=\cos
U^3,\; N^{3}_4=k\sin U^2\sin U^3,\; N^{3}_5 =-k\cos U^5\sin U^6,\;
N^{3}_6=k\cos U^6$, where  $k= const,$ and the other  functions
$M^{a}_{a '}, N^{a}_{a'}$ vanish. Substituting these functions
$M^{a}_{a '}, N^{a}_{a '},$ into Eqs.(\ref{m53}), we obtain the Lax
representation of the Euler-Lagrange system for Lagrangian
$$L= \sum_{\beta_1 =1}^3{U^{\beta_1}_xU^{\beta_1}_y} + 2\cos U^3U^1_yU^2_x +
\sum_{\beta_2 =4}^6{U^{\beta_2}_xU^{\beta_2}_y}$$
\begin{equation}
\label{m54} +\cos U^6(U^4_yU^5_x+U^4_xU^5_y) - \cos U^6(U^4_yU^5_x
- U^4_xU^5_y) + 2k \cos U^3 \cos U^6.
\end{equation}

\vspace*{3mm} \noindent {\bf Example 4.}(An integrable extension of
the sine-Gordon equation)

Notice that the system of Euler-Lagrange equations for the
Lagrangian (\ref{m54}) admits  the reduction which is similar to one
shown in \cite{Get}. Indeed, inserting the expressions
$$U^1_x=-\cos U^3U^2_x,\;U^1_y =-\frac{1}{\cos
U^3}U^2_y,\;$$
$$U^4_x=-\frac{1}{\cos U^6}U^5_x,\;U^4_y=-\cos U^6U^5_y$$
in this system, we arrive at the system
$$U^2_{xy} + U^3_yU^2_x ctgU^3
+\frac{1}{\sin U^3\cos U^3}U^3_xU^2_y = 0,\; U^3_{xy} - U^2_yU^2_x
tg U^3 +k\sin U^3\cos U^6 = 0,$$
$$U^5_{xy} + U^5_yU^6_x ctgU^6
+\frac{1}{\sin U^6\cos U^6}U^5_xU^6_y = 0,\; U^6_{xy} - U^5_yU^5_x
tg U^6 +n\cos U^3\sin U^6 = 0.$$ The latter system is not the
system of Euler-Lagrange equations. However, it admits the
Backlund transformation defined by the equalities
$$U^2_y=\frac{\cos U^3}{1+\cos U^3}V^1_y,\;\;U^2_x=\frac{1}{1+\cos
U^3}V^1_x, \;\;U^3=V^2,$$
$$U^5_y=\frac{1}{1+\cos U^6}V^3_y,\;\;U^5_x=\frac{\cos U^6}{1+\cos
U^6}V^3_x,\;\;U^6=V^4,$$ to the Euler-Lagrange one
$$ V^1_{xy}+\frac{1}{\sin V^2}V^1_yV^2_x+\frac{1}{\sin
V^2}V^2_yV^1_x=0,$$
$$V^2_{xy}-\frac{\sin V^2}{(1+\cos V^2)^2}V^1_yV^1_x+k\sin V^2\cos
V^4=0,
$$
\begin{equation}
\label{m55} V^3_{xy}+\frac{1}{\sin V^4}V^3_yV^4_x+\frac{1}{\sin
V^4}V^4_yV^3_x=0,
\end{equation}
$$V^4_{xy}-\frac{\sin V^4}{(1+\cos V^4)^2}V^3_yV^3_x+
k\cos V^2\sin V^4=0.$$ The Lagrangian for the system (\ref{m55})
is of the form
\begin{equation}
\label{m56} L=V^1_xV^1_y tg^2\frac{V^2}{2}+V^2_xV^2_y+V^3_xV^3_y
tg^2\frac{V^4}{2}+V^4_xV^4_y+2k\cos V^2\cos V^4.
\end{equation}

\noindent {\bf Remark 1.} The system (\ref{m55}) could be considered
as a "doubled" complex sine-Gordon system. Indeed, inserting $V^1 =
V^3, V^2 = V^4$ in (\ref{m55}), we arrive at the complex sine-Gordon
system. Further, assuming $V^1 = V^3=V,\; V^2 = V^4=0$ we obtain
well known sine-Gordon equation.

\vspace*{3mm}

\noindent {\bf Remark 2.} The metric associated with the Lagrangian
(\ref{m56}) is the product of two black hole metrics \cite{WitBH}.

\vspace*{3mm}

\noindent {\bf Remark 3.} The system (\ref{m55}) admits the Lax
representation which can be written in the form (\ref{m51}), where
$$ \theta^1_2=\frac{1}{2\cos \frac{V^2}{2}}(V^1_xdx+V^1_ydy),
\; \theta^1_3= -\frac{\sin \frac{V^2}{2}}{2\cos^2
\frac{V^2}{2}}(V^1_xdx-V^1_ydy),$$
$$\theta^2_3=\frac{1}{2}(V^2_xdx-V^2_ydy),\;
\theta^4_5=\frac{1}{2\cos \frac{V^4}{2}}(V^3_xdx+V^3_ydy),$$
$$\theta^4_6=
-\frac{\sin \frac{V^4}{2}}{2\cos^2
\frac{V^4}{2}}(V^3_xdx-V^3_ydy),\;
\theta^5_6=\frac{1}{2}(V^4_xdx-V^4_ydy),$$
$$ \omega^{a}_{a'}=\lambda
M^{a}_{a'}dx+\frac{1}{\lambda}N^{a}_{a'}dy,$$
$$
M^{a}_4=M^1_{a'}=0,\;N^{a}_4=N^1_{a'}=0,\;\;
$$
$$ M^2_5=\sin\frac{V^2}{2}\sin\frac{V^4}{2},\;
M^2_6=-\sin\frac{V^2}{2}\cos\frac{V^4}{2},$$
$$M^3_5=-\cos\frac{V^2}{2}\sin\frac{V^4}{2},\;\;
M^3_6=\cos\frac{V^2}{2}\cos\frac{V^4}{2},$$
$$ N^2_5=k\sin\frac{V^2}{2}\sin\frac{V^4}{2},\;\;
N^2_6=k\sin\frac{V^2}{2}\cos\frac{V^4}{2},$$
$$ N^3_5=k\cos\frac{V^2}{2}\sin\frac{V^4}{2},\;\;
N^3_6=n\cos\frac{V^2}{2}\cos\frac{V^4}{2}.$$

\vspace*{3mm} Notice, that we construct the Lax representation of
the system (\ref{m55}) with the help of  the structure equations of
the symmetric space $SO(6)/(SO(3)\times SO(3))$. Further we give
examples of new integrable systems similar to Example 4. These
systems differ from ones mentioned in theorem 5 but the Lax
representations for them are associated with the symmetric spaces of
the form $G/(H_1 \times H_2)$.

\begin{theorem}
To every symmetric space of the form $SO(p+3)/(SO(p)\times
SO(3))\;(p\geq 3)$ there  corresponds the Lax representation of Euler's
equations for Lagrangian
$$L=g_{\alpha_1 \beta_1}U^{\alpha_1}_xU^{\beta_1}_y +
a_{\alpha_1 \beta_1}U^{\alpha_1}_xU^{\beta_1}_y-
2(p-2)[V^{1}_xV^{1}_ytg^2\frac{V^{2}}{2}+V^{2}_xV^{2}_y]$$
\begin{equation}
\label{m57}
  +4(p-2)M^{b}_{b ' }N^{b '}_{b}.
\end{equation}
Here $g_{\alpha_1 \beta_1},a_{\alpha_1 \beta_1}$ are of the same
sense as in previous theorems, Latin indices $a ,b $ range from 1 to
$p$; indices $a ', b '$ range from $p+1$ to $p+3$ and functions
$M^{b}_{b '}, N^{b}_{b'}$ will be defined in the proof.
\end{theorem}
{\bf Proof.} We use the embedding  of the group $SO(p+3)$ into
$GL(p+3)$ and rewrite the structure equations of the symmetric space
$SO(p+3)/(SO(p)\times SO(3))$ in form (\ref{m51}), where indices  $a
,b =\overline{1,p},\; a ',b' =\overline{p+1,p+3}.$ Let the
coefficients $H^{a}_{b \delta_1}=-H^{b}_{a \delta_1}$ be defined by
the embedding of the Lie algebra  $\mathfrak{so}(p)$ into the
$\mathfrak{gl}(p)$. Then these coefficients satisfy the identities:
\begin{equation}
\label{10c} H^{a}_{b \delta_1}C^{\delta_1}_{\beta_1 \gamma_1} =
H^{a}_{d [\gamma_1}H^{d}_{|b |\beta_1 ]},\; \;(p-2)H^{a}_{b
\beta_1}H^{b}_{a \gamma_1} =h_{\beta_1 \gamma_1},
\end{equation}
where $C^{\delta_1}_{\beta_1 \gamma_1}$ and $h_{\beta_1 \gamma_1}$
are the structure constants and the Killing metric of
$\mathfrak{so}(p)$ whit respect to the basis
$\Phi^{\alpha_1}=T^{\alpha_1}_{\beta_1} dU^{\beta_1}$.

Assume
\begin{equation}
\label{m59}
 \theta^{a}_{b}=
H^{a}_{b \beta_1}T^{\beta_1}_{\gamma_1}U^{\gamma_1}_xdx,\;\;
\theta^{p+1}_{p+2}=\frac{1}{2\cos
\frac{V^{2}}{2}}(V^{1}_xdx+V^{1}_ydy),
\end{equation}
\begin{equation}
\label{m60}
 \theta^{p+1}_{p+3}=
-\frac{\sin\frac{V^{2}}{2}}{2\cos^2\frac{V^2}{2}}
(V^1_xdx-V^1_ydy),\;\;
\theta^{p+2}_{p+3}=\frac{1}{2}(V^2_xdx-V^2_ydy),
\end{equation}
$$ \omega^{a}_{a'}=\lambda
M^{a}_{a'}dx+ \frac{1}{\lambda}N^{a}_{a'}dy$$ in (\ref{m51}). Then
we obtain the completely integrable systems for the functions
 $M^{a}_{a '}, N^{a}_{a '}$:
$$
M^{a}_{a',\beta_1}=0,\; M^{a}_{p+1,V^1}=\frac{1}{2\cos
\frac{V^{2}}{2}} M^{a}_{p+2}+\frac{\sin \frac{V^{2}}{2}}{2\cos^2
\frac{V^{2}}{2}} M^{a}_{p+3}, \;M^{a}_{p+1,V^2}=0,
$$
$$
M^{a}_{p+2,V^1}=-\frac{1}{2\cos \frac{V^{2}}{2}} M^{a}_{p+1},\;
M^{a}_{p+2,V^2}=-\frac{1}{2}M^{a}_{p+3},\;
M^{a}_{p+3,V^1}=-\frac{\sin \frac{V^{2}}{2}}{2\cos^2
\frac{V^{2}}{2}} M^{a}_{p+1},\;
$$
$ M^{a}_{p+3,V^2}=\frac{1}{2}M^{a}_{p+2},\; N^{a}_{a',\beta_1}=
2H^{a}_{b \gamma_1}T^{\gamma_1}_{\beta_1}N^{b}_{a'},\;
N^{a}_{p+1,V^1}=\frac{1}{2\cos \frac{V^{2}}{2}}
N^{a}_{p+2}-\frac{\sin \frac{V^{2}}{2}}{2\cos^2 \frac{V^{2}}{2}}
N^{a}_{p+3}, $
$$
N^{a}_{p+1,V^2}=0,\; N^{a}_{p+2,V^1}= -\frac{1}{2\cos
\frac{V^{2}}{2}} N^{a}_{p+1},\;
N^{a}_{p+2,V^2}=\frac{1}{2}N^{a}_{p+3},
$$
$$
N^{a}_{p+3,V^1}=\frac{\sin \frac{V^{2}}{2}}{2\cos^2 \frac{V^{2}}{2}}
N^{a}_{p+1},\; N^{a}_{p+3,V^2}=-\frac{1}{2}N^{a}_{p+2}.
$$
One can verify that integrating these systems and then substituting
the found solutions and expressions (\ref{m59}),(\ref{m60}) into
Eqs.(\ref{m51}), we obtain the Lax representation of the
Euler-Lagrange system for Lagrangian (\ref{m57}). $\square$

\begin{theorem}
To every symmetric space

$SO(p+2)/(SO(p)\times SO(2))\;(p\geq 3)$ there corresponds the Lax
representation of Euler's equations for Lagrangian
$$L=[g_{\alpha_1 \beta_1}(U^{\gamma_1})+
a_{\alpha_1
\beta_1}(U^{\gamma_1})]U^{\alpha_1}_{x}U^{\beta_1}_{y}-2(p-2)k^2V_xV_y+
4(p-2)M^{b}_{b'}N^{b'}_{b},$$ where $k=const.$
\end{theorem}

{\bf Proof.} The Lax representation  of these system we obtain
substituting
\begin{equation}
\label{m61} \theta^{a}_{b}=H^{a}_{b \beta_1}
T^{\beta_1}_{\gamma_1}U^{\gamma_1}_xdx,\;\theta^{p+1}_{p+2}=kV_ydy,
\end{equation}
$$ \omega^{a}_{a'}=\lambda M^{a}_{a'}dx+
\frac{1}{\lambda}N^{a}_{a'}dy$$ in (\ref{m51}). Here the functions
$M^{a}_{a '}, N^{a}_{a '}$ satisfy the following equations:
\begin{equation}
\label{m62} M^{a}_{a', \beta_1}=0,\;M^{a}_{p+1,V}=kM^{a}_{p+2},\;
M^{a}_{p+2,V}=-kM^{a}_{p+1},
\end{equation}
$$N^{a}_{a',\beta_1}=H^{a}_{b \gamma_1}T^{\gamma_1}_{\beta_1}N^{b}_{a'},\;
N^{a}_{a' ,V}=0.$$$\square$

\vspace*{3mm} \noindent {\bf Example 5.}(The system associated with
$SO(5)/(SO(3)\times SO(2)).$)

Choose the local coordinates and left-invariant forms on the group
$SO(3)$ as in Example 1. Assume $M^1_5=\sin U^2\sin U^3,\;
M^2_5=-\cos U^2\sin U^3,\; M^3_5=\cos U^3,\;M^{a}_4=0,\;
N^1_{a'}=N^2_{a} =0,\; N^3_4=l \sin kV,\; N^3_5=l\cos
kV,\;l=const,\;k=const$. Then we obtain the Lax representation of
the Euler-Lagrange system for the Lagrangian
$$L=\sum_{\alpha_1=1}^{3}U^{\alpha_1}_xU^{\alpha_1}_y+k^2V_xV_y+2\cos U^3
U^1_yU^2_x+2l\cos U^3\cos kV.$$

Carrying out the reductions and the Backlund transformation
analogous to Example 1, we obtain the system
$$V^1_{xy}+\frac{1}{\sin V^2}V^1_yV^2_x+\frac{1}{\sin V^2}
V^2_yV^1_x=0,$$
\begin{equation}
\label{m63} V^2_{xy}-\frac{\sin V^2}{(1+\cos
V^2)^2}V^1_yV^1_x+l\sin V^2\sin V^3=0,
\end{equation}
$$V^3_{xy}+\frac{l}{k}\cos V^2\sin kV^3=0$$
which is the Euler's system for the Lagrangian
\begin{equation}
\label{m64} L=V^1_xV^1_y
tg^2\frac{V^2}{2}+V^2_xV^2_y+k^2V^3_xV^3_y + 2l\cos V^2\cos kV^3.
\end{equation}

This system admits the Lax representation of the form (\ref{m51}),
where
$$ \theta^1_2=\frac{1}{2\cos \frac{V^2}{2}}(V^1_xdx+V^1_ydy),\;
\theta^1_3= -\frac{\sin \frac{V^2}{2}}{2\cos^2
\frac{V^2}{2}}(V^1_xdx-V^1_ydy),$$
$$\theta^2_3=\frac{1}{2}(V^2_xdx-V^2_ydy),\;
\theta^4_5=kV^3_xdx,$$
$$ \omega^{A}_{A'}=\lambda
M^{A}_{A'}dx+\frac{1}{\lambda}N^{A}_{A'}dy,$$
$M^{A}_4=M^1_5=N^1_4=N^1_5=0,\; M^2_5=-\sin \frac{V^2}{2}, \;
M^3_5=\cos \frac{V^2}{2},\; N^2_4=l\sin \frac{V^2}{2}\sin kV^4,\;
N^2_5=l\sin \frac{V^2}{2}\cos kV^3,\; N^3_4=l\cos
\frac{V^2}{2}\sin kV^3,\; N^3_5=l\cos \frac{V^2}{2}\cos kV^3.$

Notice that the integrable systems with Lagrangian
$$L=V^1_xV^1_y
tg^2\frac{V^2}{2}+V^2_xV^2_y+k^2V^3_xV^3_y + Q$$ were completely
studied by Meshkov A.G. and Demskoi D.K. (see e.g.
\cite{Mesh},\cite{DM}).

\section{Acknowledgments}
The authors are grateful to Ferapontov E.V., Stepanov N.A. and Tuzov
Y.V. for helpful discussions.

\bibliographystyle{amsplain}

\end{document}